\documentstyle[12pt]{article}

\date{}

\def\pd{\partial}

\def\m{\mu}
\def\n{\nu}
\def\a{\alpha}

\def\b{\beta}
\def\g{\gamma}
\def\G{\Gamma}
\def\d{\delta}
\def\r{\rho}

\def\s{\sigma}
\def\S{\Sigma}
\def\na{\nabla}

\def\l{\lambda}
\def\te{\theta}
\def\na{\bigtriangledown}
\def\p{\phi}

\def\hR{\hat R}

\def\ep{\epsilon}
\def\cc{{\cal C}}

\begin{document}

\begin{titlepage}

\begin{center}
\hfill hep-th/0007134\\
\hfill UM--TH/00-02

\vskip 1.5 cm
{\Large \bf Ricci-Parallelizable Spaces in the NS-NS Sector:} 
\vskip .7 cm
{\Large \bf $AdS_3\times S^7$}

\vskip 1.5 cm

{\large Leopoldo A. Pando Zayas}\\

\vskip 1cm
Randall Laboratory of Physics\\
The University of Michigan\\
Ann Arbor, Michigan 48109-1120\\
\vskip .8cm

{\tt lpandoz@umich.edu}

\end{center}

\vskip 1.8 cm

\begin{abstract}
We provide a class of nondilatonic solutions to the NS-NS sector of string
theory. The solutions consist of products of Ricci-parallelizable
spaces with adjusted radii. A representative of this class, 
$AdS_3\times S^7$, is presented in detail. Some comments on 
possible brane connections are made.
\end{abstract}
\end{titlepage}
            
\section{Introduction}
Central to the current nonperturbative understanding of M-theory has
been the prominent role of certain supergravity solutions. Brane
solutions were crucial in uncovering dualities between different string
theories. The D-brane counterpart of these supergravity solutions helped
achieve progress in certain supersymmetric gauge theories. Another
class of supergravity solutions that has played an important role has
the form $AdS_d\times S^{D-d}$, where $D=10, 11$. These solutions are
central in the AdS/CFT correspondence. A particularly attractive feature
of a subclass of these solutions is the fact that the dilaton is
constant, since this allows for far more control in the CFT side.

Two solutions that have been missing in the class of nondilatonic
$AdS_d\times S^{D-d}$ were the cases $d=3$ and $d=7$. Here we provide
such solutions as representatives of a class of solutions we
outlined. The relevance of these nondilatonic AdS backgrounds to the
AdS/CFT remains to be determined but it seems sensible to expect that
these solutions will play an important role.

A natural generalization of
the AdS/CFT correspondence consists of replacing $S^d$ by appropriate
spaces that are less symmetric than spheres,  the 
typical example
is replacing $S^5$ by $T^{1,1}$ \cite{klebanov}.  This generalization
allows for a richer structure in the CFT side. Some of the solutions
belonging to the class we discuss in this paper allow for naturally
substituting the sphere for less symmetric manifolds.

The fact that parallelizable spaces have a distinguished position in the
context of string theory has been known since the 80's when most
of the sigma model analysis took place \cite{callan,tseytlin}. It was
realized then that parallelizable spaces satisfy the conformal
invariance equations. There was, however, a negative feature
associated with the fact that in general their central charge is not
vanishing. The fact that the central charge is proportional to the
scalar curvature imposes a very restrictive condition for critical
models,  $R=0$. This
condition will naively lead us to physically uninteresting situations.

In this paper we show how to enforce $R=0$ in a class of physically
interesting solutions. Basically, we consider 10D spaces that are
products of parallelizable spaces. This ansatz provides a solution to
the NS-NS sector of string theory provided we adjust the radii of the
two manifolds involved to enforce $R=0$. A representative of this class,
$AdS_3\times S^7$, is treated in detail in section 3 after 
outlining the general argument in section 2. Section 4 contains some
comments on possible brane interpretations of solutions containing
$AdS_3$ as a factor in the near-horizon geometry. Conclusions are 
drawn in section 5.

\section{Parallelizability in the NS-NS sector}

Consider the NS-NS sector of string theory given by the following 
action\footnote{A more fundamental view of this analysis can be
presented from the sigma-model point of view \cite{callan,
tseytlin}. Here, however, we are going to concentrate on the spacetime
counterpart assuming $D=10$.}

\begin{equation}
S={1\over 2\kappa_{10}^2}\int d^{10}x\sqrt{-g}e^{-2\p}\left
( R+4(\partial \p)^2-{1\over 12}H^2\right).
\end{equation}

We are going to specialize to the case of $\p={\mbox const.}$
The equation of motions are

\begin{eqnarray}
\pd_M(\sqrt{-g}H^{MRS})&=&0, \nonumber \\
R_{MN}&=&{1\over 4}H_{MPQ}H_N^{PQ}, \nonumber \\
R&=&{1\over 12}H^2
\end{eqnarray}
Here the third line is a constraint coming from the dilaton
equation. Noting that the second line implies $R=H^2/4$, one sees that the
only solution is $R=H^2=0$.

To make the contact with parallelizable spaces more evident, recall that
the generalized Riemann tensor is the Riemann tensor calculated from
the generalized connection
$\hat \G^\l{}_{\m\n}=\G^\l{}_{\m\n}-{1\over 2}H^\l{}_{\m\n}$,

\begin{eqnarray}
\hR_{\a\b\g\r}&=&R_{\a\b\g\r}+{1\over 2}\na_\g H_{\a\b\r}-{1\over 2}\na_\r
H_{\a\b\g}+{1\over 4}H_{\s\a\g}H^{\s}{}_{\r\b}
-{1\over 4}H_{\s\a\r}H^{\s}{}_{\g\b}, \nonumber \\
\hR_{\a\b}&=&R_{\a\b}-{1\over 4}H_{\a\g\r}H_{\b}{}^{\g\r}
+{1\over 2}\na^{\g}H_{\g\a\b}
\end{eqnarray}
A space is parallelizable if $\hR_{\a\b\g\r}=0$ and Ricci-parallelizable
if $\hR_{\a\b}=0$. 
The second equation makes the relevance to the equations of
motion of the NS-NS sector of string theory with a constant dilaton
explicit. We now have that the Einstein equation is nothing but imposing
the vanishing of the
symmetric part of the generalized Ricci tensor and that the equation
of motion for the NS-NS two-form field is nothing but the vanishing of
the antisymmetric part. This
means that any Ricci-parallelizable space furnishes us with a potential
solution. One needs, however, to take into consideration the constraint
coming from the dilaton equation of motion $R=0$. For this purpose we consider
a 10D space that is the direct product of two spaces, in other words,
consider splitting the index $M$ into $(\m,m)$ with $\m=0,\ldots,d$
and $m=d+1, \ldots, 9-d$ with $x^M=(x^\m,y^m)$ and the following ansatz 
for the metric and
antisymmetric tensors

\begin{eqnarray}
g_{\m n}&=&0, \quad g_{\m\n}=g_{\m\n}(x), \quad g_{mn}= g_{mn}(y), \nonumber \\
B_{\m n}&=&0, \quad B_{\m\n}=B_{\m\n}(x), \quad B_{m n}=B_{mn}(y).
\end{eqnarray}
Under these assumptions, all we have to do is to adjust by hand the
``radii'' of the subspaces to guarantee that the total scalar curvature
is zero. The situation is strikingly similar to the Freund-Rubin
compactification
\cite{fr}.  Here we have that any pair of {\it Ricci-parallelizable}
spaces with opposite scalar curvature  provides us with a solution. Note
that in the FR compactification all one needs is a pair of Einstein
spaces with opposite scalar curvature. Actually, the solutions presented
here are more similar to Englert's generalization \cite{englert} of the
FR compactification  in the sense  that the antisymmetric tensor is
nonvanishing in both subspaces. It is precisely in this sense that the
construction presented here generalizes part of that of \cite{duff}. The
$AdS_7\times S^3$ background that can be constructed based on the scheme
described above has nontrivial three-form tensor in both factors of the
space.

\section{Examples}
The first, now standard,  result on parallelizable manifolds was
obtained by Cartan and
Schouten \cite{cartan} (see also \cite{braaten} for a modern
discussion). It states that only group manifolds and $S^7$ admit an 
{\it absolute parallelism}, i.e., are globally 
parallelizable in a way that leaves the geodesics of the manifold
unaltered (with a totally antisymmetric torsion). This classic work was
generalized in \cite{wolf} to pseudo-Riemannian spaces. A particularly
interesting generalization to homogeneous spaces was obtained in the
series 
\cite{singhof} \footnote{D. L\"ust explicitly constructed the
parallelizing torsion for a class of homogeneous coset spaces
\cite{lust}. Other explicit examples have been considered recently in
\cite{gavrilik,castellani}}.  An earlier discussion of these homogeneous spaces
was conducted in \cite{wolf1,manturov} from a different point of
view. For a partial list of parallelizable spaces the reader is referred
to \cite{wolf1,manturov}\footnote{Most of the classification has been
carried out under restrictive conditions, such as, for example, assuming
that $G$ is simple. The general classification for any $G$ is a more
difficult question which depends more intricately on concrete embeddings
of $H$ into $G$ \cite{lust, gavrilik,castellani,castellani1}.}. Some common 
representatives are
group manifolds, Stiefel manifolds with the exception of
spheres,\footnote{$S^1$, $S^3$ and $S^7$ are the only parallelizable
spheres. This fact is related to the corresponding division algebras and
to the Hopf fiberings.} and $G/T$ where $T$ is a non-maximal toral
subgroup.

A proposition  by Wolf \cite{wolf} clarifies why parallelizable spaces are of
such interest in various areas of mathematics. It states that for $M$ a 
connected differentiable manifold,
there are natural one-one correspondences between (i) absolute
parallelisms $\phi$ on $M$;
\footnote{This is a statement about vector fields, usually phrased  as
spaces that can be nonsingularly combed.}
 (ii) smooth trivializations of the frame
bundle $B \to M$; (iii) smooth connections $\G$ on $B \to M$ with
holonomy group reduced to the identity. This proposition shows that the
problem can be tackled using methods of differential geometry, algebraic
topology or representation theory.

It is worth noting that any parallelizable space is necessarily
Ricci-parallelizable but the converse need not be true. Much of the
literature, both in physics and mathematics, has concentrated on
parallelizable spaces but we will
keep in mind that the condition for being a solution is weaker. Two
examples of spaces that are Ricci-parallelizable but are not
parallelizable are the squashed $S^7$ and a particular embedding of
$(SU(2)\times SU(2))/U(1)$.

It is also possible to consider more than two factors. The scheme
outlined above  works perfectly well for the product of any number of
Ricci-parallelizable spaces. Recalling that the only parallelizable
spheres are $S^1$, $S^3$ and $S^7$ one could form:
$AdS_3\times S^3\times S^3\times S^1$ or for that matter
$AdS_3\times S^3\times T^4$, where $T^4$ is trivially parallelizable and
$AdS_3\times S^3\times K3$, where $K3$ is Ricci flat and therefore Ricci
parallelizable with trivial torsion. The
latter model is related to the D1/D5 system and has received much
attention recently.

\subsection{$AdS_3\times S^7$}

To make the above description  more precise let us consider the following
background for the $AdS_3$ part

\begin{eqnarray}
ds^2&=&R_1^2\left({du^2\over u^2}+u^2(-dt^2+dx^2)\right), \nonumber \\
B_{tx}&=&R_1^2u^2.
\end{eqnarray}
Some properties of this background are

\begin{equation}
R_{ab}=-{2\over R_1^2}g_{ab}, \quad R=-{6\over R_1^2}, \quad H_{utx}=2R_1^2u.
\end{equation}
Note that the parallelizing torsion is proportional to the product of
the natural dreibein $R_1^{-1}(R_1/u)(R_1u)^2$. This fact is not very
revealing at this point since this is basically the only three-form in
$AdS_3$, but in $S^7$ the situation will be much different.  One can
check, using these explicit relations, that it satisfies the equations of
motion.

The seven sphere and its parallelizing torsion has received a great deal of
attention in the 11D supergravity context. Englert \cite{englert} used 
the dual of
the parallelizing torsion to generalized the FR solution. The
relation of this parallelizing torsion to the octonions was made
explicit, for example,  in \cite{gursey}. Here we are going to
restrict ourselves to the simplest case of the round $S^7$. For a
more detailed analysis, including the squashed $S^7$, the reader
should consult \cite{englert1}; for the physical relevance of
torsion in the supergravity context see \cite{duff1}. The
following analysis follows most closely references \cite{englert1,duff2,duff3}.
We consider the following metric on $S^7$

\begin{eqnarray}
ds^2&=&R_2^2\left( d\m^2+{1\over 4}\sin^2\m(\s_i-\S_i)^2
+(\cos^2{\m\over 2}\s_i+\sin^2{\m\over 2}\S_i)^2\right), \nonumber \\
\s_1&=&\cos\psi_1d\te_1+\sin\psi_1\sin\te_1d\p_1, \nonumber \\
\s_2&=&-\sin\psi_1d\te_1+\cos\psi_1\sin\te_1d\p_1, \nonumber \\
\s_3&=&d\psi_1+\cos\te_1d\p_1,
\end{eqnarray}
similar relations define $\S_i$ but with the subindex $1$ replaced by $2$.
Following \cite{englert1,duff2,duff3}, one naturally introduces the following
Siebenbein
\begin{equation}
e^0=R_2d\m, \quad e^i={R_2\over 2}\sin\m(\s_i-\S_i), \quad
e^{\hat i}=R_2 (\cos^2{\m\over 2}\s_i+\sin^2{\m\over 2}\S_i),
\end{equation}
where $i,\, {\hat i}\,=\, 1,2,3$. In this orthonormal frame the parallelizing
torsion is simply proportional to the octonionic multiplication table. It
will be more suggestive to call the values of the parallelizing three-form in
the orthonormal frame the octonionic structure constant. This is simply the
result of Cartan-Schouten \cite{cartan} that is discussed in \cite{braaten}:
the parallelizing torsion in the orthonormal frame is given by the structure
constant.

For imaginary octonions one has
\begin{equation}
O_aO_b=-\d_{ab}+f_{abc}O_c, \quad f_{0i\hat j}=-\d_{ij},
\quad f_{ij\hat k}=-\ep_{ijk}, \quad f_{\hat i\hat j \hat k}=\ep_{ijk}
\end{equation}
More precisely one has $H_{abc}= R_2^{-1}f_{abc}$. In the orthonormal
frame we obtain 

\begin{equation}
R_{ab}={3\over 2R_2^2}\d_{ab}, \quad H_{ab}^2={6\over R_2^2}\d_{ab},
\end{equation}
from which  the equations of motions  follow automatically.  Our last task
is to find the relation between the radii of $AdS_3$ and $S^7$ that
makes the total scalar curvature vanish,

\begin{equation}
R=R_{AdS_3}+R_{S^7}={-6\over R_1^2}+{21\over 2R_2^2}=0 
\Rightarrow {R_1\over R_2}={2\over \sqrt{7}}.
\end{equation}

It is worth stressing a rather unique property of this solution.
The background we are considering is an
exact string solution in the sense that it  does not receive $\a'$
corrections. The $AdS_3$ part is simply a WZW model on $SL(2,R)$ and
although $S^7$ is not a group manifold, there is a CFT structure defined
on it \cite{s7cft} that parallels the WZW construction. This
construction  is very similar to the WZW model
in the sense that the energy-momentum tensor is given by the Sugawara
construction using the currents that generate the associated Kac-Moody
algebra. The central charge of this CFT is $c=7k/(k+12)$ where $k$ is
the level of the KM algebra. In the semiclassical
approximation that we discussed here, $k=1/\a'\to \infty$ and $c=7$.

\section{Relation to Branes}
One question that naturally arises is what is the relation of this class
of solutions to strings (1-branes). Although a general analysis is 
possible we will
concentrate on backgrounds possibly having $AdS_3$ as a factor in the
near-horizon geometry. For that
purpose we consider the following ansatz

\begin{eqnarray}
ds^2&=&e^{2A(r)}(-dt^2+dx^2)+e^{2B(r)}dr^2+e^{2C(r)}g_{mn}dy^mdy^n,
\nonumber \\
B_{tx}&=&e^{C_1(r)}, \quad B_{mn}=e^{C_2(r)}b_{mn},\quad  \phi=\mbox{const.},
\end{eqnarray}
where $b_{mn}$ generates the parallelizing torsion $h_{mnp}$ on a
seven-dimensional manifold  with metric $g_{mn}$. At this point, even
without further calculation, we draw certain conclusions about the
strings and their near-horizon geometry. Note that if we want a ``round''
$AdS_3$ in the near horizon limit, $A(r)$ must be related to $B(r)$ in an
obvious way to give the same radius for the whole $AdS_3$. On the other
hand,  for the typical brane ansatz for the transverse part one has
$C(r)=B(r)+\ln r$. This means that the near-horizon geometry of the  standard 
string can be reached only for a 7D having the same radius
as $AdS_3$. More explicitly, in the near-horizon limit to have a
``round'' $AdS_3$ we need $A(r)=\ln (r R_1)$ and $B(r)=\ln (R_1/r)$; for the
standard string $B(r)=C(r)-\ln r=\ln (R_2/r)$. This means that we can only 
achieve both
conditions (round $AdS_3$ and standard string ansatz) when the
radii of the two parallelizable manifolds are the same ($R_1=R_2$). This
is possible for $AdS_3\times S^3\times {\cal N}$ with ${\cal N}$ being
$T^4$ or
$K3$  which is dictated by the ``effective''
dimensionality of the parallelizable spaces. This simple analysis
also shows that $AdS_3\times S^7$ can not be reached from a standard
string $(C(r)=B(r)+\ln r$). 
The ingredients for the equations of motion are

\begin{eqnarray}
0&=&\pd_r\left(e^{-2A-B+7C}\cc_1\right), \nonumber \\
0&=&\pd_r\left(e^{2A-B+3C}\cc_2\right), \nonumber \\
R_{\m\n}&=&-\eta_{\m\n}e^{2(A-B)}(A''+2(A')^2-A'B'), \nonumber \\
{1\over 4}H_{\m\n}^2&=&-{1\over 2}\eta_{\m\n}\cc_1^2e^{-2A-2B}, \nonumber \\
R_{rr}&=&-2(A''+(A')^2-A'B')-7(C''+(C')^2-B'C'), \nonumber \\
{1\over 4}H_{rr}^2&=&-{1\over 2}\cc_1^2e^{-4A}+{1\over 2}\cc_2^2e^{-4C}
b^2, \nonumber \\
R_{mn}&=&R_{mn}(g)-g_{mn}e^{2(C-B)}(C''+7(C')^2-B'C'), \nonumber \\
{1\over 4}H_{mn}^2&=&{1\over 4}h_{mn}^2e^{2C_2-4C}+ 
{1\over 2}\cc_2^2b_{mn}^2e^{-2B-2C},
\end{eqnarray}
where $\cc_i=(e^{C_i})'$, $b^2=b_{mn}b_{pq}g^{mp}g^{nq}$,
$b_{mn}^2=b_{mp}b_{nq}g^{pq}$. If we want to engineer a solution for
which the Einstein equation for $(n,m)$ is of the form
$R_{mn}=h_{mn}^2/4$, we must take $C_2=2C$. In order to get rid of the
term containing $b_{mn}^2$ we need $\cc_2=0$. Altogether this implies
that $C_2$ and $C$ are constants; the term containing $g_{mn}$ also
vanishes for constant $C$. 

For constant $C_2$ and $C$, the whole system can be uniquely solved with
$A'={q\over 2}e^B$. This uniquely fixes the metric to be $AdS_3$. Namely, going
to a new coordinate $dR=e^Bdr$ the 3D part of the metric can be written
as 

\begin{equation}
ds^3=e^{qR}(-dt^2+dx^2)+dR^2,
\end{equation}
which is nothing but $AdS_3$ in Poincare coordinates. We conclude that,
if one insists in a parallelizable manifold as the near-horizon limit of
a string, the only solution is $AdS_3$. It might be natural to expect that $AdS_3 \times S^7$ arises as the limit of branes intersections. A very exhaustive study of brane configurations and the corresponding $AdS$ factor was carried out in \cite{kostas}. This analysis reveals that the $AdS_3\times S^7$  geometry does not arise from brane intersections either \cite{kostas}.

It is worth noting that actually for most of the explicit constructions
of parallelizable manifolds, the parallelizing torsion in the
orthonormal frame is constant. This means that the two-form that
generates this torsion should be proportional to the product of
vielbeins. This might conspire to produce $b_{mn}^2$ proportional to
$g_{mn}$ and therefore allow for non constant $C$. Here we will not
explore such possibility as it is clear that it depends on the explicit
form of $b_{mn}$ but we think it is a very feasible scenario for
constructing branes.  Another  very viable possibility \footnote{This is
the generalization needed for the D1/D5 system} is to generalize 
the  ansatz we considered here to allow $\p=\p(r)$ with the 
condition $\phi=const$
enforced only in the near-horizon limit.

\section{Conclusions}
We have described a class of solutions to the NS-NS sector of string
theory with constant dilaton. Some of the representatives of this class
were known from supergravity analysis \cite{castellani}, but some are
new. This class of solutions puts a number of
particular cases under one unifying scheme, in particular we now
recognize the relation between $AdS_3\times S^3\times S^3\times S^1$, 
$AdS_3\times S^3\times T^4$, $AdS_3\times S^3\times K3$ and $AdS_3\times S^7$.

We have described in detail the case of
$AdS_3\times S^7$. We have also discussed the possibility of
constructing strings whose near-horizon geometry is of the form $AdS_3
\times {\cal N}$ where ${\cal N}$ is a Ricci-parallelizable manifold. We
found certain restrictions on ${\cal N}$ for a specific brane ansatz. In
particular we showed that there is no brane solution if one insists that
the Einstein equation of motion of  ${\cal N}$ is strictly the
condition of Ricci-parallelizability. We pointed out, however, that it
might be possible, by a more detailed analysis, to construct string
solutions with the desired near-horizon geometry.

One of the interesting characteristics of this class of solutions is
associated with the fact that the NS-NS sector is a universal sector of
various string theories. Using U-dualities one could generate
backgrounds with R-R fields. This universality property has been
exploited  in  
\cite{tsensns},
through the use of a web of dualities, to generate new string 
solutions whose brane
worldvolume is a curved space. 

We would like to point out that a general scheme for
studying the supersymmetric properties of these solutions is lacking and
one would have to deal with it  on a case by case basis. In other words,
Ricci-parallelizability does not guarantee neither does it forbid 
preservation of some fraction of the supersymmetry. For example, in
the case of $AdS_3\times {\cal N}$ it has been established
\cite{castellani} that in the framework of type IIB some solutions are 
supersymmetric but some are not.  Another feature pointing to the need
for a more particular analysis is that  the supersymmetry tranformation
of these solutions are 
model-depending. As we have pointed out, these solutions can be
embedded in N=2 or N=1 supersymmetric string theories as well as 11D
supergravity. This and other related matters will be further discussed
in a future publication.

\begin{center}
{\large \bf Acknowledgments}
\end{center}
I  am grateful to  M. Cederwall for several explanations on topics related
to $S^7$; M. Duff for encouragement and many useful suggestions;
M. Einhorn for encouragement and helpful criticism. I would especially like to thank A.A. Tseytlin for
discussions on very related issues, advice and pointing out an error in the previous version.  I would
like to acknowledge helpful suggestions by J.T. Liu, J.X. Lu
and S. Monni. I  would also like to acknowledge  the Office of the
Provost at the University of Michigan and the High Energy Physics 
Division of the Department of Energy for support.

\end{document}